\documentclass[aip,reprint,author-year,nofootinbib]{revtex4-1}

\usepackage{graphicx}
\usepackage{amsmath,amssymb}             
\usepackage{stmaryrd}					
\usepackage[normalem]{ulem}
\usepackage{color}
\usepackage[dvipsnames]{xcolor}
\usepackage{color,hyperref}
\definecolor{darkblue}{rgb}{0.0,0.0,0.4}
\definecolor{red}{rgb}{0.7,0.0,0.0}
\definecolor{green}{rgb}{0.0,0.5,0.0}
\hypersetup{colorlinks,breaklinks,
            linkcolor=darkblue,urlcolor=darkblue,
            anchorcolor=darkblue,citecolor=RoyalBlue}

\def\mnras{Mon. Not. R. Astron. Soc.}

\def\jcap{Journal of Cosmology and Astroparticle Physics}

\defcitealias{Jasche2015BORGSDSS}{\textsc{borg sdss}}

\begin{document}


\title{Cosmic web-type classification using decision theory}


\author{Florent Leclercq}
\email{florent.leclercq@polytechnique.org}
\affiliation{Institut d'Astrophysique de Paris (IAP), UMR 7095, CNRS -- UPMC Universit\'e Paris 6, Sorbonne Universit\'es, 98bis boulevard Arago, F-75014 Paris, France}
\affiliation{Institut Lagrange de Paris (ILP), Sorbonne Universit\'es,\\ 98bis boulevard Arago, F-75014 Paris, France}
\affiliation{\'Ecole polytechnique ParisTech,\\ Route de Saclay, F-91128 Palaiseau, France}

\author{Jens Jasche}
\affiliation{Excellence Cluster Universe, Technische Universit\"at M\"unchen,\\ Boltzmannstrasse 2, D-85748 Garching, Germany}

\author{Benjamin Wandelt}
\affiliation{Institut d'Astrophysique de Paris (IAP), UMR 7095, CNRS -- UPMC Universit\'e Paris 6, Sorbonne Universit\'es, 98bis boulevard Arago, F-75014 Paris, France}
\affiliation{Institut Lagrange de Paris (ILP), Sorbonne Universit\'es,\\ 98bis boulevard Arago, F-75014 Paris, France}
\affiliation{Department of Physics, University of Illinois at Urbana-Champaign,\\ 1110 West Green Street, Urbana, IL~61801, USA}
\affiliation{Department of Astronomy, University of Illinois at Urbana-Champaign,\\ 1002 West Green Street, Urbana, IL~61801, USA}


\date{\today}

\begin{abstract}
\noindent We propose a decision criterion for segmenting the cosmic web into different structure types (voids, sheets, filaments, and clusters) on the basis of their respective probabilities and the strength of data constraints. Our approach is inspired by an analysis of games of chance where the gambler only plays if a positive expected net gain can be achieved based on some degree of privileged information. The result is a general solution for classification problems in the face of uncertainty, including the option of not committing to a class for a candidate object. As an illustration, we produce high-resolution maps of web-type constituents in the nearby Universe as probed by the Sloan Digital Sky Survey main galaxy sample. Other possible applications include the selection and labelling of objects in catalogues derived from astronomical survey data.
\end{abstract}


\maketitle



\section{Introduction}

Building accurate maps of the cosmic web from galaxy surveys is one of the most challenging tasks in modern cosmology. Rapid progress in this field took place in the last few years with the introduction of inference techniques based on Bayesian probability theory \citep{Kitaura2009,Jasche2010a,Nuza2014,Jasche2015BORGSDSS}. This facilitates the connection between the properties of the cosmic web, thoroughly analyzed in simulations \citep[e.g.][]{Hahn2007a,Aragon-Calvo2010,Cautun2014}, and observations \citep[see][for a review on the interface between theory and data in cosmology]{Leclercq2014Varenna}.

In \citet{Leclercq2015ST}, we conducted a fully probabilistic analysis of structure types in the cosmic web as probed by the Sloan Digital Sky Survey (SDSS) main galaxy sample. This study capitalized on the large-scale structure inference performed by \citet{Jasche2015BORGSDSS} using the \textsc{borg} \citep[Bayesian Origin Reconstruction from Galaxies,][]{Jasche2013BORG} algorithm. As the full gravitational model of structure formation \textsc{cola} \citep[COmoving Lagrangian Acceleration,][]{Tassev2013} was used, our approach resulted in the first probabilistic and time-dependent classification of cosmic environments at non-linear scales in physical realizations of the large-scale structure conducted with real data. Using the \citet{Hahn2007a} definition \citep[see also its extensions,][]{Forero-Romero2009,Hoffman2012}, we obtained three-dimensional, time-dependent maps of the posterior probability for each voxel to belong to a void, sheet, filament or cluster.

These posterior probabilities represent all the available structure type information in the observational data assuming the framework of $\Lambda\mathrm{CDM}$ cosmology. Since the large-scale structure cannot be uniquely determined from observations, uncertainty remains about how to assign each voxel to a particular structure type. The question we address in this letter is how to proceed from the posterior probabilities to a particular choice of assigning a structure type to each voxel. Decision theory \citep[see, for example,][]{Berger1985} offers a way forward, since it addresses the general problem of how to choose between different actions under uncertainty. A key ingredient beyond the posterior is the utility function that assigns a quantitative profit to different actions for all possible outcomes of the uncertain quantity. The optimal decision is that which maximizes the expected utility.

After setting up the problem using our example and briefly recalling the relevant notions of Bayesian decision theory, we will discuss different utility functions and explore the results based on a particular choice.

\section{Method}
\label{sec:Decision theory Method}

The decision problem for structure-type classification can be stated as follows. We have four different web-types that constitute the ``space of input features:'' \{$\mathrm{T}_0=$ void, $\mathrm{T}_1=$ sheet, $\mathrm{T}_2=$ filament, $\mathrm{T}_3=$ cluster\}. We want to either choose one of them, or remain undecided if the data constraints are not sufficient. Therefore our ``space of actions'' consists of five different elements: \{$a_0=$ ``decide void,'' $a_1=$ ``decide sheet,'' $a_2=$ ``decide filament,'' $a_3=$ ``decide cluster,'' and $a_{-1}=$ ``do not decide.''\} The goal is to write down a decision rule prescribing which action to take based on the posterior information.

Bayesian decision theory states that the action $a_j$ that should be taken is that which maximizes the expected utility function (conditional on the data $d$), given in this example by
\begin{equation}
\label{eq:utility_function}
U(a_j(\vec{x}_k)|d) = \sum_{i=0}^3 G(a_j|\mathrm{T}_i) \, \mathcal{P}(\mathrm{T}_i(\vec{x}_k)|d) ,
\end{equation}
where $\vec{x}_k$ labels one voxel of the considered domain, $\mathcal{P}(\mathrm{T}_i(\vec{x}_k)|d)$ are the posterior probabilities of the different structure types given the data, and $G(a_j|\mathrm{T}_i)$ are the gain functions that state the  profitability of each action, given the ``true'' underlying structure. Formally, $G$ is a mapping from the space of input features to the space of actions. For our particular problem, it can be thought of as a $5\times4$ matrix $\mathbf{G}$ such that $\mathbf{G}_{ij} \equiv G(a_j|\mathrm{T}_i)$, in which case eq. \eqref{eq:utility_function} can be rewritten as a linear algebra equation, $\mathbf{U}~=~\mathbf{G}.\mathbf{P}$ where the 5-vector $\mathbf{U}$ and the 4-vector $\mathbf{P}$ contain the elements $\mathbf{U}_j \equiv U(a_j(\vec{x}_k)|d)$ and $\mathbf{P}_i \equiv \mathcal{P}(\mathrm{T}_i(\vec{x}_k)|d)$, respectively.

Let us consider the choice of gain functions. Several choices are possible. For example, the 0/1-gain functions reward a correct decision with 1 for each voxel, while an incorrect decision yields 0. This leads to choosing the structure type with the highest posterior probability. While this seems like a reasonable choice, we need to consider that a decision is taken in each voxel, whereas we are interested in identifying structures as objects that are made of many voxels. For instance, since clusters are far smaller than voids, the \textit{a priori} probability for a voxel to belong to a cluster is much smaller than for the same voxel to belong to a void. To treat different structures on an equal footing, it makes sense to reward the correct choice of structure type $\mathrm{T}_i$ by an amount inversely proportional to the average volume $V_i$ of one such structure. In the following, we use the prior probability as a proxy for the volume fractions,
\begin{equation}
\mathcal{P}(\mathrm{T}_i) \approx \frac{V_i}{V_0+V_1+V_2+V_3} .
\end{equation}
We further introduce an overall cost for choosing a structure with respect to remaining undecided, leading to the following specification of the utility,
\begin{equation}
\label{eq:gain_functions}
G(a_j|\mathrm{T}_i) = \left\{
\begin{array}{ll}
      \dfrac{1}{\mathcal{P}(\mathrm{T}_i)} - \alpha & \mathrm{if~} j \in \llbracket0,3\rrbracket \mathrm{~and~} i=j, \\
      -\alpha & \mathrm{if~} j \in \llbracket0,3\rrbracket \mathrm{~and~} i \ne j, \\
      0 & \mathrm{if~} j=-1.\\
\end{array} 
\right.
\end{equation}
This choice limits 20 free functions to only one free parameter, $\alpha$. With this set of gain functions, making (or not) a decision between structure types can be thought of as choosing to play or not to play a gambling game costing $\alpha$. Not playing the game, i.e. remaining undecided ($j =-1$), is always free ($G(a_{-1}|\mathrm{T}_i) = 0$ for all $i$). If the gambler decides to play the game, i.e. to make a decision ($j \in \llbracket0,3\rrbracket$), they pay $\alpha$ but may win a reward, $\frac{1}{\mathcal{P}(\mathrm{T}_i)}$, by betting on the correct underlying structure ($i=j$).

In the absence of data, the posterior probabilities in equation \eqref{eq:utility_function} are the prior probabilities $\mathcal{P}(\mathrm{T}_i)$, which are independent of the position $\vec{x}_k$, and the utility functions are, for $j \in \llbracket 0,3 \rrbracket$,
\begin{eqnarray}
U(a_j) & = & \sum_{i=0}^3 G(a_j|\mathrm{T}_i) \, \mathcal{P}(\mathrm{T}_i) \nonumber \\
& = & \left(\dfrac{1}{\mathcal{P}(\mathrm{T}_j)} - \alpha\right) \mathcal{P}(\mathrm{T}_j) - \sum_{\substack{i=0 \\ i\neq j}}^3 \alpha \, \mathcal{P}(\mathrm{T}_i) \nonumber \\
& = & 1 - \alpha \left( \mathcal{P}(\mathrm{T}_j) + \sum_{\substack{i=0 \\ i\neq j}}^3 \mathcal{P}(\mathrm{T}_i) \right) \nonumber \\
\label{eq:utility_function_prior_1}
& = & 1 - \alpha, \\
\mathrm{and} \quad U(a_{-1}) &=& 0.
\label{eq:utility_function_prior_2}
\end{eqnarray}
Equations \eqref{eq:utility_function_prior_1} and \eqref{eq:utility_function_prior_2} mean that, in the absence of data, this reduces to the roulette game utility function, where, if correctly guessed, \textit{a priori} unlikely outcomes receive a higher reward, inversely proportional to the fraction of the probability space they occupy. Betting on outcomes according to the prior probability while paying $\alpha=1$ leads to a \textit{fair game} with zero expected net gain. The gambler will always choose to play if the cost per game is $\alpha \leq 1$ and will never play if $\alpha > 1$.

The posterior probabilities update the prior information in light of the data, providing an advantage to the gambler through privileged information about the outcome. In the presence of informative data, betting on outcomes based on the posterior probabilities will therefore ensure a positive expected net gain and the gambler will choose to play even if $\alpha>1$. Increasing the parameter $\alpha$ therefore represents a growing \textit{aversion for risk} and limits the probability of losing. Indeed, for high $\alpha$, the gambler will only play in cases where the posterior probabilities give sufficient confidence that the game will be won, i.e. that the decision will be correct.

\section{Maps of structure types in the SDSS}
\label{sec:Decision theory Maps}

\begin{figure*}
\begin{center}
\includegraphics[width=\textwidth]{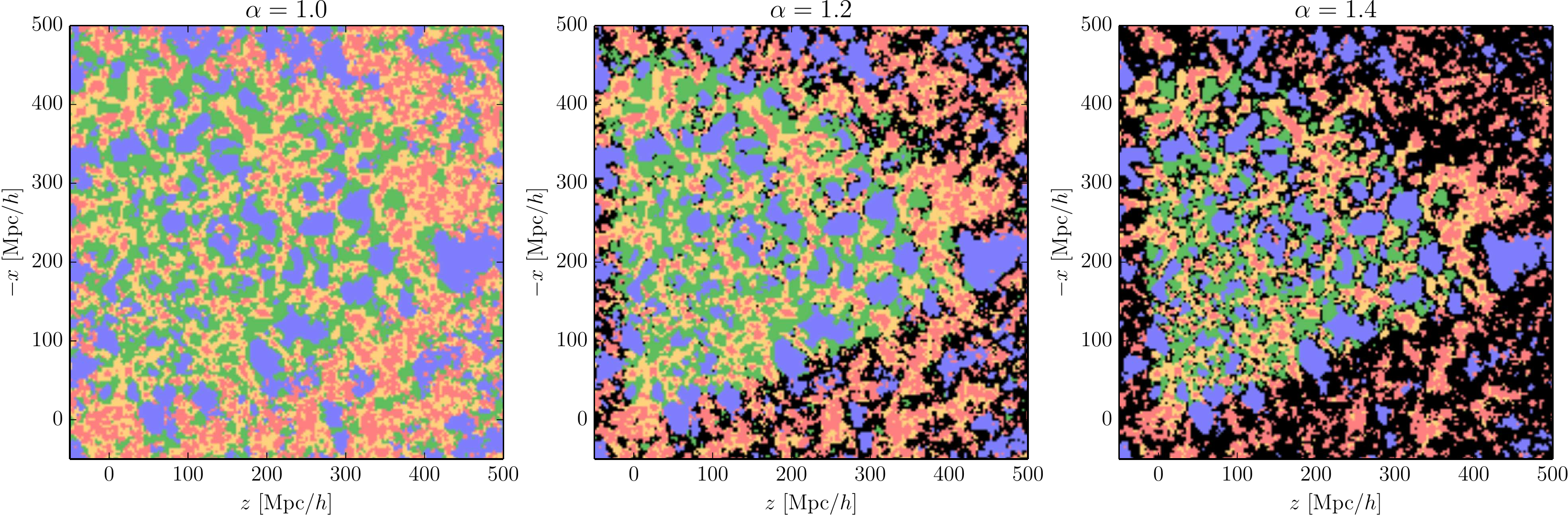}\\
\vspace{3pt}
\includegraphics[width=\textwidth]{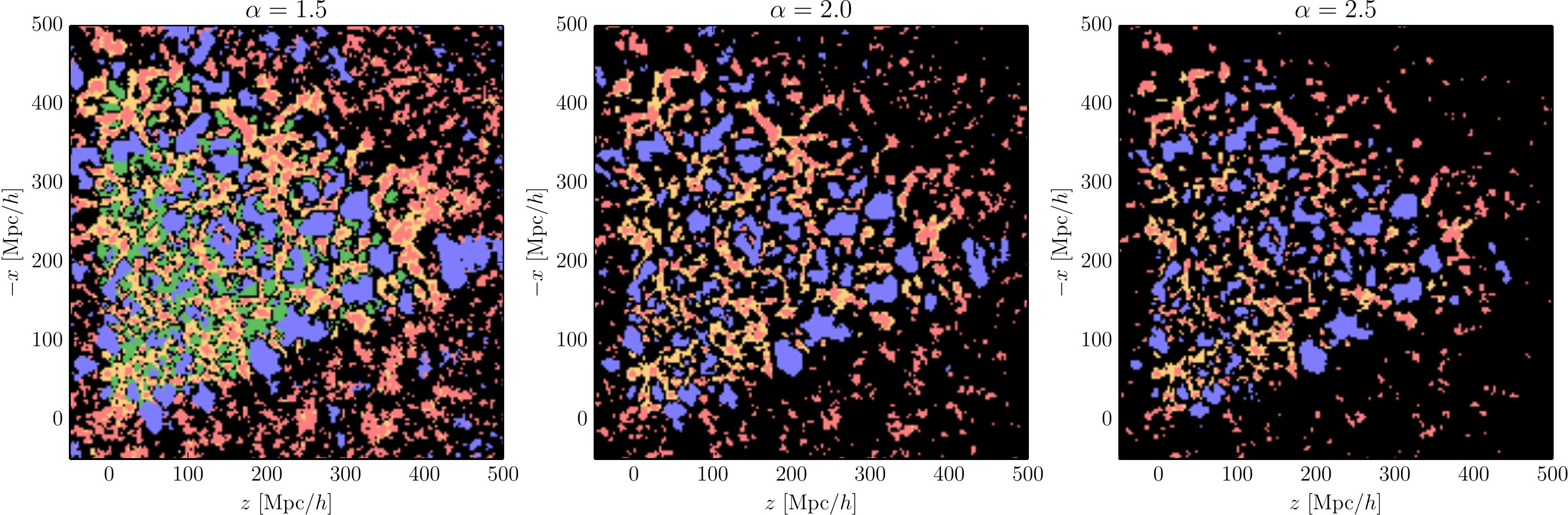}\\
\vspace{3pt}
\includegraphics[width=\textwidth]{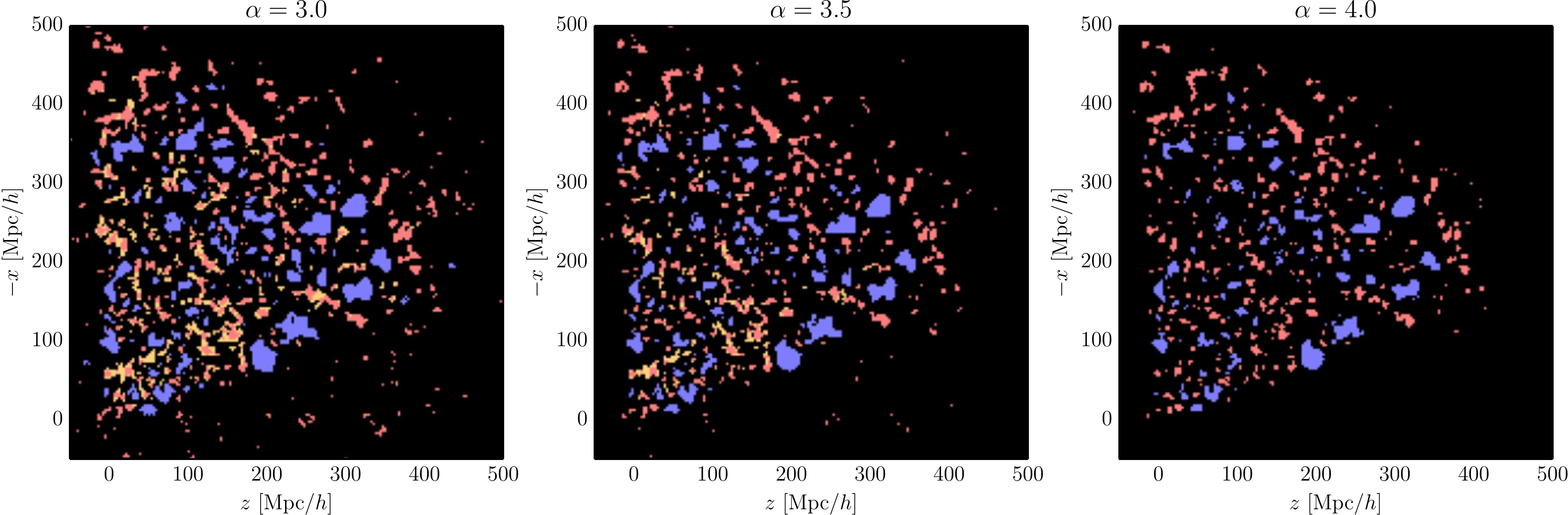}
\caption{Slices through maps of structure types in the late-time large-scale structure, at $a=1$. The colour coding is blue for voids, green for sheets, yellow for filaments, and red for clusters. Black corresponds to regions where data constraints are insufficient to make a decision. The parameter $\alpha$, defined by equation \eqref{eq:gain_functions}, quantifies to the risk aversion in the map: $\alpha=1.0$ corresponds to the most speculative map of the large-scale structure, and maps with $\alpha \geq 1$ are increasingly conservative. These maps are based on the posterior probabilities inferred by \citet{Leclercq2015ST} and on the Bayesian decision rule subject of the present work.\label{fig:decision_map_final}}
\end{center}
\end{figure*}

\begin{figure*}
\begin{center}
\includegraphics[width=\textwidth]{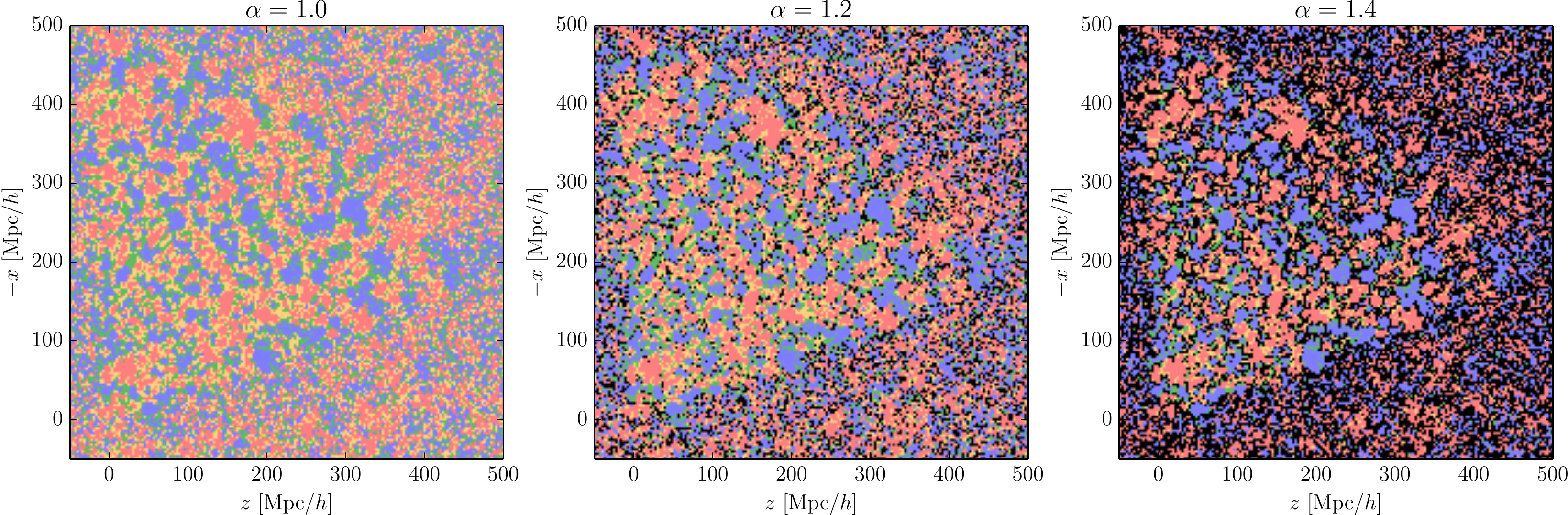}\\
\vspace{3pt}
\includegraphics[width=\textwidth]{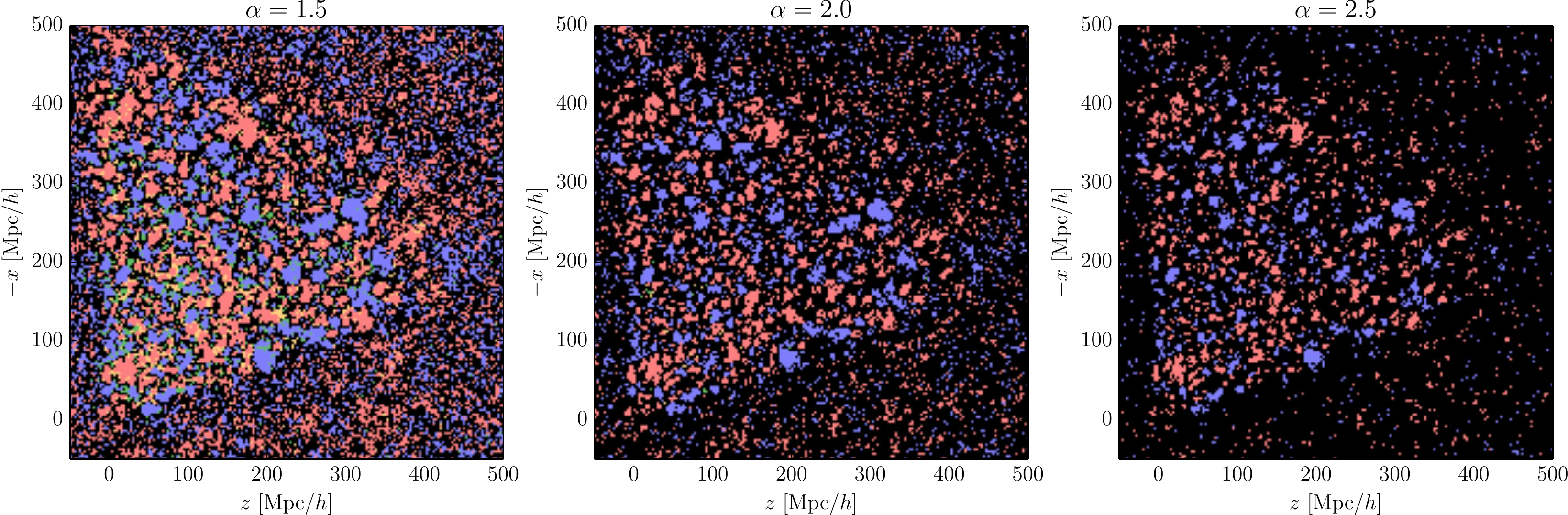}\\
\vspace{3pt}
\includegraphics[width=\textwidth]{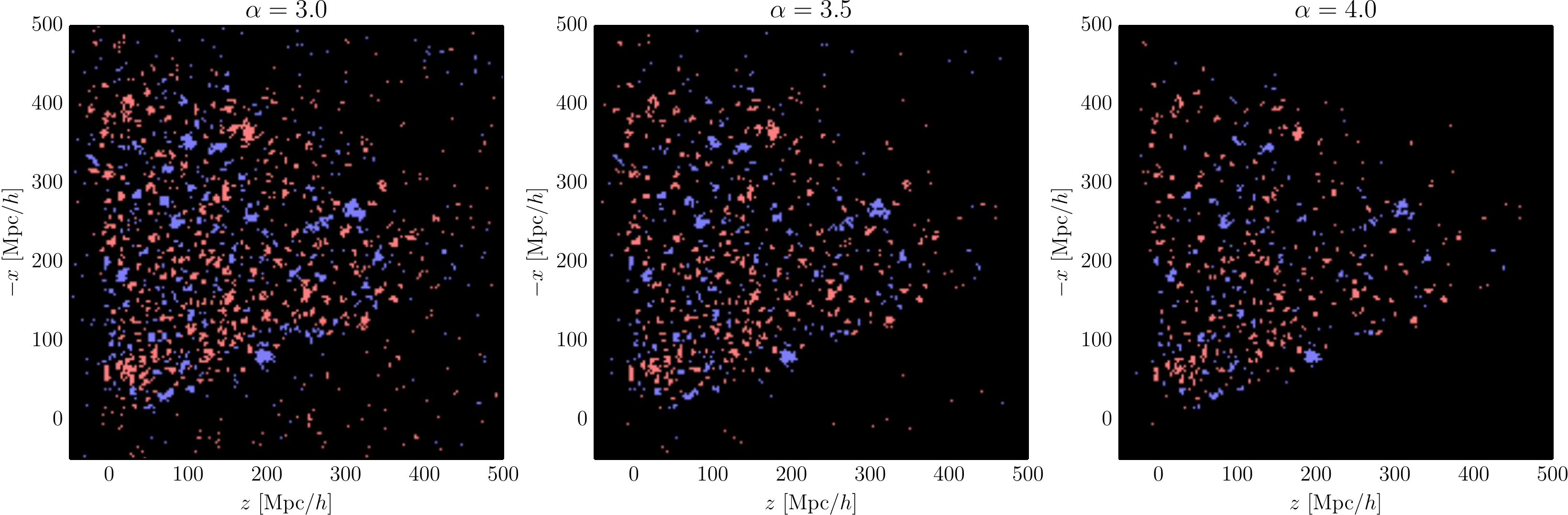}
\caption{Same as figure \ref{fig:decision_map_final} for the primordial large-scale structure, at $a=10^{-3}$.
\label{fig:decision_map_initial}}
\end{center}
\end{figure*}

We applied the above decision rule to the web-type posterior probabilities presented in \citet{Leclercq2015ST}, for different values of $\alpha \ge 1$ as defined by equation \eqref{eq:gain_functions}. In doing so, we produced various maps of the volume of interest, consisting of the northern Galactic cap of the SDSS main galaxy sample and its surroundings. Slices through these three-dimensional maps\footnote{For all slices shown, we kept the coordinate system of \citet{Jasche2015BORGSDSS}.} are shown in figure \ref{fig:decision_map_final} for the late-time large-scale structure (at $a=1$) and in figure \ref{fig:decision_map_initial} for the primordial large-scale structure (at $a=10^{-3}$).

When the game is fair (namely when $\alpha =1$), it is always played, i.e. a decision between one of the four structure types is always made. This results in the \textit{speculative map} of structure types (top left panel of figures \ref{fig:decision_map_final} and \ref{fig:decision_map_initial}). There, a decision is made even in regions that are not constrained by the data (at high redshift or outside of the survey boundaries), based on prior betting odds. 

By increasing the value of $\alpha >1$, we demand higher confidence in making the correct decision. This yields increasingly \textit{conservative maps} of the Sloan volume (see figures \ref{fig:decision_map_final} and \ref{fig:decision_map_initial}). In particular, at high values of $\alpha$, the algorithm makes decisions in the regions where data constraints are strong \citep[see figures 3 and 6 in][]{Leclercq2015ST}, but often stays undecided in the unobserved regions. It can be observed that even at very high values, $\alpha \gtrsim 3$, a decision for one structure is made in some unconstrained voxels (typically in favour of the structure for which the reward is the highest: clusters in the final conditions, and clusters or voids in the initial conditions). This effect is caused by the limited number of samples used in our analysis. Indeed, because of the finite length of the Markov Chain, the sampled representation of the posterior has not yet fully converged to the true posterior. For this reason, the numerical representation of the posterior can be artificially displaced too much from the prior, which results in an incorrect web-type decision. This effect could be mitigated by obtaining more samples in the original \textsc{borg} analysis (for an increased computational cost); or can be avoided by further increasing $\alpha$, at the expense of also degrading the map in the observed regions. We found the value of $\alpha=4$ (bottom right panel of figures \ref{fig:decision_map_final} and \ref{fig:decision_map_initial}) to be the best compromise between reducing the number of unobserved voxels in which a decision is made to a tiny fraction and keeping information in the volume covered by the data.

As expected, structures for which the prior probabilities are the highest disappear first from the map when one increases $\alpha$: betting on these structures being poorly rewarded, this choice is avoided in case of high risk aversion. In the final conditions (figure \ref{fig:decision_map_final}), we found that sheets completely disappear for $\alpha \approx 1.68$ and filaments for $\alpha \approx 4.01$. In the initial conditions (figure \ref{fig:decision_map_initial}), the critical value is around $\alpha \approx 2.36$ for both sheets and filaments. In the most conservative maps displayed in figures \ref{fig:decision_map_final} and \ref{fig:decision_map_initial} ($\alpha = 4.0$), the SDSS data provide extremely high evidence for the voids and clusters shown. In constrained parts, extended regions belonging to a given structure type may not have the expected shape. This is true in particular for filamentary regions. Several factors can explain this: first, slicing through filaments make them appear as dots; second, with the dynamic \citet{Hahn2007a} definition, filament regions often extend out into sheets and voids, and their static skeleton geometry is not the most prominent at the voxel scale (3 Mpc/$h$ in this work).

As detailed in \citet{Jasche2015BORGSDSS}, data constraints are propagated by the structure formation model assumed in the inference process (second-order Lagrangian perturbation theory) and therefore radiate out of the SDSS boundaries. For this reason, for moderate values of $\alpha$, web-type classification can be extended beyond the survey boundaries to regions influenced by data. This can be observed in figures \ref{fig:decision_map_final} and \ref{fig:decision_map_initial}, where one can see, for instance, that the shape of voids that intersect the mask is correctly recovered. Similarly, the classification of high-redshift structures confirms that the treatment of selection effects by \textsc{borg} is correctly propagated to web-type analysis.

We finally comment on the required computational resources for the complete chain for running \textsc{borg}, computing the web-type posterior, and making a decision. Inference with \textsc{borg} is the most expensive part: on average, one sample is generated in 1500 seconds on 16 cores \citep{Jasche2015BORGSDSS}. Then, in each sample, tidal shear analysis \citep{Leclercq2015ST} is a matter of a few seconds. Once the web-type posterior is known, making a decision, which is the subject of the present letter, is almost instantaneous. Therefore, once the density field has been inferred, which is useful for a much larger variety of applications, our method is substantially cheaper than several state-of-the-art techniques for cosmic web analysis \citep[e.g. the method of ][for detecting filaments]{Tempel2013,Tempel2014}.

\section{Conclusions}
\label{sec:Conclusions}

In this letter, we proposed a rule for optimal decision making in the context of cosmic web classification. We described the problem set-up in Bayesian decision theory and proposed a set of gain functions that permit an interpretation of the problem in the context of game theory. This framework enables the dissection of the cosmic web into different elements (voids, sheets, filaments, and clusters) given their prior and posterior probabilities and naturally accounts for the strength of data constraints.

As an illustration, we produced three-dimensional templates of structure types with various risk aversion, describing a volume covered by the SDSS main galaxy sample and its surrounding. These maps constitute an efficient statistical summary of the inference results presented in \citet{Leclercq2015ST} for cross-use with other astrophysical and cosmological data sets.

Beyond this specific application, our approach is more generally relevant to the solution of classification problems in the face of uncertainty. For example, the construction of catalogues from astronomical surveys is directly analogous to the problem we describe here: it simultaneously involves a decision about whether or not to include a candidate object and which class label (e.g.\ star or galaxy) to assign to it.

\acknowledgments

FL acknowledges funding from an AMX grant (\'Ecole polytechnique ParisTech). JJ is partially supported by a Feodor Lynen Fellowship by the Alexander von Humboldt foundation. BW acknowledges funding from an ANR Chaire d'Excellence (ANR-10-CEXC-004-01) and the UPMC Chaire Internationale in Theoretical Cosmology. This work has been done within the Labex \href{http://ilp.upmc.fr/}{Institut Lagrange de Paris} (reference ANR-10-LABX-63) part of the Idex SUPER, and received financial state aid managed by the Agence Nationale de la Recherche, as part of the programme Investissements d'avenir under the reference ANR-11-IDEX-0004-02. This research was supported by the DFG cluster of excellence ``\href{www.universe-cluster.de}{Origin and Structure of the Universe}''.

\bibliography{structuretypes}

\end{document}